\documentclass{jaa}
\DeclareRobustCommand{\VAN}[3]{#2}
\let\VANthebibliography\thebibliography
\def\thebibliography{\DeclareRobustCommand{\VAN}[3]{##3}\VANthebibliography}

\usepackage{graphicx}
\usepackage[flushleft]{threeparttable}
\begin{document}\sloppy

\title{A New Measurement of the Spin and Orbital Parameters of the High Mass X-ray Binary Centaurus X-3 using AstroSat}

\author{Parisee Shirke\textsuperscript{*}, Suman Bala\textsuperscript{}, Jayashree Roy\textsuperscript{} and Dipankar Bhattacharya\textsuperscript{}}
\affilOne{\textsuperscript{}Inter-University Centre for Astronomy and Astrophysics (IUCAA), Post Bag 4, Ganeshkhind, Pune 411 007, India\\}

\twocolumn[{

\maketitle

\corres{parisee@iucaa.in}

\msinfo{30 October 2020}{6 January 2021}

\begin{abstract}
We present the timing results of out-of-eclipse observations of Centaurus X-3 spanning half a binary orbit, performed on 12-13 December, 2016 with the Large Area X-ray Proportional Counter (LAXPC) on-board AstroSat. The pulse profile was confirmed to exhibit a prominent pulse peak with a secondary inter-pulse. The systemic spin period of the pulsar was found to be $4.80188 \pm 0.000085$ s in agreement with its spin up trend. The spin up timescale seems to have increased to $7709 \pm 58$ yr that points to negative torque effects in the inner accretion disk. We also report the derived values of projected semi-major axis and orbital velocity of the neutron star.
\end{abstract}


}]


\doinum{12.3456/s78910-011-012-3}
\artcitid{\#\#\#\#}
\volnum{000}
\year{0000}
\pgrange{1--}
\setcounter{page}{1}
\lp{1}

\section{Introduction}

The X-ray source Centaurus X-3 was detected by Chodil {\em et al.} (1967) in a rocket experiment. From the X-ray and optical studies, it has been established that it is an eclipsing High Mass X-ray Binary (HMXB) comprising of an X-ray pulsar with a spin period of $\sim$ 4.8 s and an
optical companion star with an orbital period of $\sim$ 2.1 days (Giacconi {\em et al.} 1971; Schreier {\em et al.} 1972b). The optical companion is an O6.5 II-III supergiant known as Krzeminski’s star (Verbunt \& van den Heuvel 1995; Schreier {\em et al.} 1972a; Krzeminski 1974).

The X-ray pulsar has a mass of $1.49 \pm 0.08 \text{ M}_{\odot}$ (Rawls {\em et al.} 2011) and the optical counterpart has a mass of $ 20.5 \pm 0.7 \text{ M}_{\odot}$ (Hutchings {\em et al.} 1979; Ash {\em et al.} 1999) and radius of $11.4 \pm 0.7 \text{ R}_{\odot}$ (Falanga {\em et al.} 2015). The eclipse lasts for $\sim 20\%$ of the orbit (Nagase 1989). The latest estimate of the distance to Centaurus X-3 is $5.7 \pm 1.5$ kpc (Thompson \& Rothschild 2009). 

The high, persistent luminosity of the X-ray pulsar $L_{2-10 \text{ keV}} \sim 5 {\times} 10^{37} \text{ ergs/s}$ is sustained by mass transfer arising from a combination of predominantly disk accretion and an excited stellar wind (Petterson 1978; Bildsten {\em et al.} 1997). The stellar wind seems to be driven thermally from the X-ray heated face of the atmosphere of the Krzeminski’s star (Day \& Stevens 1993). This is supported by the observation of ellipsoidal variations in the optical light curve produced by the tidally deformed supergiant (Tjemkes {\em et al.} 1986) indicative of a star filling its Roche lobe. Centaurus X-3 is one of the very few X-ray binaries that exhibit a secular decay of the orbital period, which could be attributed to tidal dissipation (Burderi {\em et al.} 2000). The mass transfer causes a secular spin up trend with wavy fluctuations over several years. The resultant tidal interaction between a distorted supergiant and its companion neutron star results in an orbital decay characterised by $|\dot{P_b}/{P_b}| \sim 1.8 {\times} 10^{-6} \text{ yr}^{-1}$ (Kelley {\em et al.} 1983; Nagase {\em et al.} 1992).

The out-of-eclipse phase-averaged spectrum in 1-40 keV band is usually modelled by a power law with a high energy cut-off (Suchy {\em et al.} 2008) and a Gaussian iron line along with a soft X-ray excess below $5 \text{ keV}$ (White \& Swank 1982). The soft excess seems to arise due to the scattering of X-rays by ambient wind around an obscuring gas stream (Nagase {\em et al.} 1992). Burderi {\em et al.} (2000) interpreted it as black-body radiation with $\text{kT} \sim 0.1 \text{ keV}$. A cyclotron resonance scattering feature (CRSF) around 30 keV was first confirmed by Santangelo {\em et al.} (1988), who estimated the magnetic field to be $\sim (2.4 - 3) \times 10^{12}$ Gauss. Pulse phase-resolved spectroscopy reveals asymmetric variation of the magnetic cyclotron line energy which could be due to an offset of the dipole with respect to the neutron star center (Burderi {\em et al.} 2000).

Burderi {\em et al.} (2000) undertook broad-band (0.1-100) spectral and timing analysis of out-of-eclipse Centaurus X-3 using observations by \textit{BeppoSAX} in the year 1997. Suchy {\em et al.} (2008) presented a detailed analysis of PCA-RXTE data from observations over two consecutive binary orbits. This was followed by an in-depth pulse arrival time analysis by Raichur \& Paul (2010) from the same observation. Naik {\em et al.} (2011) carried out the spectral analysis of Centaurus X-3 using \textit{Suzaku} observations over one orbital period. 

The AstroSat mission launched in 2015 promises powerful timing and spectral capabilities for studies of compact objects (Bhattacharya 2017). Out of this class of targets, observations of neutron stars in X-ray binary systems can provide accurate measurements of orbital parameters, thus aiding the characterisation of the orbital evolution of stars in binary systems (Paul 2017). Centaurus X-3 is an example of such a system.

In this paper, we present the timing parameters of Centaurus X-3 derived from an AstroSat/LAXPC observation of 12-13 December, 2016. In Section 2, we provide details about the data and its reduction. Section 3 describes the timing analysis performed for estimation of the spin and orbital parameters. Section 4 summarises the results and discusses their physical significance. Finally, we present the conclusions in Section 5.
\section{Observations and Data Reduction}

We make use of the AstroSat/LAXPC observations from 12 December, 2016 12:31:56 (hh:mm:ss) UTC to 13 December, 2016 10:05:27 (hh:mm:ss) UTC with a useful exposure time of $53 \text{ ks}$ avoiding eclipses. This data, with Obs ID
9000000880, was retrieved from the AstroSat open public data archive\footnote{http://astrosat-ssc.iucaa.in:8080/ObservationFinder/}.

AstroSat was launched on September 28, 2015 by the Indian Space Research Organisation (ISRO) and is the first dedicated Indian astronomy mission aimed at simultaneous multi-wavelength study of astronomical sources in X-ray, optical and UV spectral bands (Agrawal 2006). It is aimed at examining their wavelength-dependent intensity variations and the underlying physical processes (Singh {\em et al.} 2014). Some of its prime scientific objectives are to understand high energy processes in X-ray binary systems and estimate magnetic fields of neutron stars.

This work makes use of Level 1 data of the Large Area X-ray Proportional Counter (LAXPC), which is a major payload on-board AstroSat with one of its primary objectives being the conduct of timing studies of X-ray binaries like Centaurus X-3 (Agrawal {\em et al.} 2017). LAXPC consists of three identical, co-aligned, independent proportional counters that register X-ray photons in the wide energy range of 3-80 keV. It has a total effective area of $\sim 6000 \text{ cm}^2$ at 10 keV, a timing precision of 10 $\mu$s and a sensitivity of 1 milliCrab in 1000 s (Antia {\em et al.} 2017; Yadav {\em et al.} 2016).
Due to a larger area (four to five times more effective area above 30 keV compared to RXTE-PCA), the sensitivity of LAXPC is unmatched in medium energy X-rays. 

The Level 1 data was converted into Level 2 data with LaxpcSoft: Format (A) software\footnote{http://astrosat-ssc.iucaa.in/?q=laxpcData \label{LaxpcSoftFormatA}} (Ver. 2018, May 21) using \texttt{laxpc\_make\_event}. We made use of the event mode data (modeEA) from the normal (default) mode of operation which is suitable for bright X-ray sources. The Level 2 data contains the arrival time, energy and identity of the detecting element (detector number and anode layer) for each X-ray photon detection event. 

A file specifying the time segments called Good Time Intervals (\textit{gti}), containing reliable data by filtering out target occultation by the Earth and passage through the South Atlantic Anomaly (SAA) regions, was produced using \texttt{laxpc\_make\_stdgti}. While the standard LAXPC software generates the \textit{gti} table based on the set criterion, some stray bad points can still remain within the \textit{gti}. These anomalies were identified by fine inspection of the light curve and manually rejected by suitably re-defining the \textit{gti}. The long observation was thus divided into 17 such time segments. 

Barycentric correction was applied to refer the photon arrival times to the barycenter of the solar system in order to correct for the relative motion of the satellite and the Earth with respect to the target source. The AstroSat Orbit File Generator Utility\footnote{http://astrosat-ssc.iucaa.in:8080/orbitgen/} was used to generate the orbit file for the observation duration. The resulting \texttt{.orb} file was provided to AstroSat's \texttt{as1bary}\footnote{http://astrosat-ssc.iucaa.in/?q=data\_and\_analysis} tool for barycentric correction after applying the Barycentric Correction Code (Ver. 2017, Jan 27). 

\begin{figure}[!t] 
    \centering
    \includegraphics[width=\columnwidth]{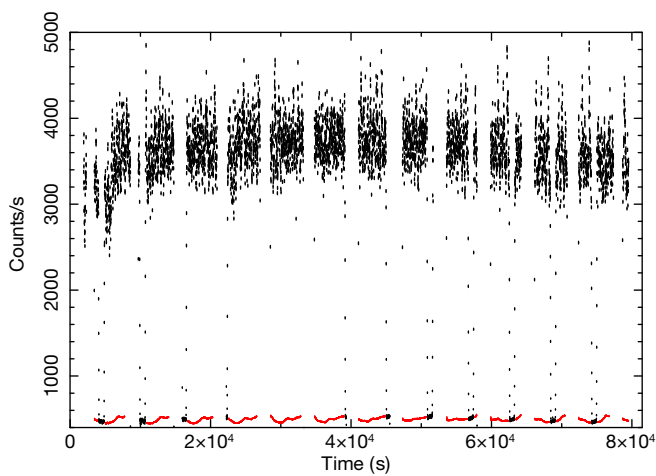}
    \caption{The complete 10s binned AstroSat/LAXPC light curve of Centaurus X-3 in 3.0-80.0 keV energy range using all the three LAXPC10, LAXPC20 and LAXPC30 detectors is presented in black colour. The gaps present in the light curve arise from the passage of AstroSat through the South Atlantic Anomaly (SAA) regions. The corresponding estimated background count rates are over-plotted in the same panel in red colour. The LAXPC error on these is $2\%$ systematic error.}
    \label{fig:lc}
\end{figure}

\section{Data Analysis}

\subsection{Timing Analysis}
The LaxpcSoft Format (A) Suite\textsuperscript{\ref{LaxpcSoftFormatA}} was used to analyse the LAXPC data. Light curves were generated using \texttt{laxpc\_make\_lightcurve} command in LaxpcSoft. The 10s binned complete light curve of the AstroSat/LAXPC observation of Centaurus X-3 on MJD 57734.57 - 57735.41 (12-13 December, 2016) combining observations of all the three LAXPC detectors, LAXPC10, LAXPC20 and LAXPC30 is depicted in Fig. \ref{fig:lc}. The corresponding estimated background rates are over-plotted in red. The systematic LAXPC error for the X-ray background is $2\%$ as per the \texttt{laxpc\_make\_backlightcurve} task for generating background light curves. Timing analysis was carried out using NASA/GSFC's HEASoft\footnote{https://heasarc.gsfc.nasa.gov/lheasoft/download.html} software package (Ver. 6.26.1 released on 2019, May 21) that comprises of FTOOLS\footnote{https://heasarc.gsfc.nasa.gov/ftools/}, a general-purpose tool to manipulate FITS files and XRONOS\footnote{https://heasarc.gsfc.nasa.gov/docs/xanadu/xronos/xronos.html} timing analysis software package (Stella \& Angelini 1992).

The eclipses of Centaurus X-3 are expected to last for $\sim 10$ h, as it has an orbital period of 2.1 days (Giacconi {\em et al.} 1971) and is known to show eclipses for $\sim 20\%$ of the orbit (Nagase 1989), during which its X-ray flux drops by an order of magnitude (Giacconi {\em et al.} 1971; Schreier {\em et al.} 1972b). The absence of such an eclipsing event during the observation time span was confirmed by using a light curve with hourly resolution (time bin size = 3600s).

The detection of periodicity in data was accomplished by performing a Fourier transform of the light curve. This was done using the \texttt{powspec} command in HEASoft that produces the power density spectrum (PDS). The 0.01s binned light curve was divided into stretches of 32768 bins per interval and results from all intervals (except the last one in which the data is insufficient to fill all the time bins) were averaged in a single frame. The resultant PDS was observed to have a sharp peak at $0.2085$ Hz. Higher harmonics were also detected upto fourth order.

\begin{figure}
    \centering
    \includegraphics[width=\columnwidth, trim={15 10 29 20 }]{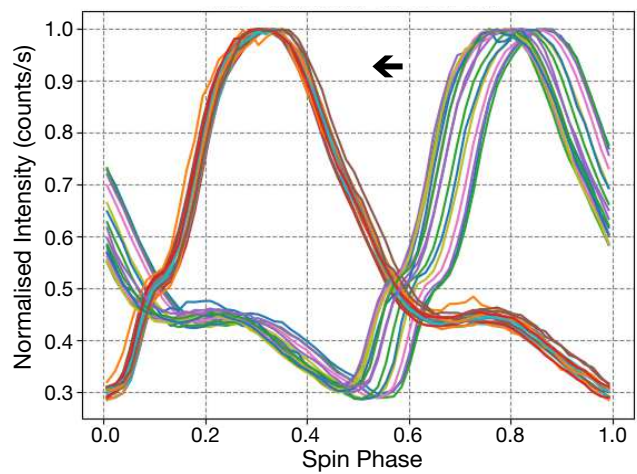}
    \caption{\small The pulse profiles, shown in different colours, obtained individually for all time segments (to the right) are overlapped on top of each other (to the left) for phase locking. The pulse profiles are averaged over the full LAXPC energy bandwidth of 3.0 to 80.0 keV.}
    \label{fig:stability}
\end{figure}

Using the reciprocal of this value as an initial approximate estimate of spin period, we searched for pulsations in the light curve with higher precision using \texttt{efsearch} in HEASoft, that determines the precise value of the local spin period $P_s$ for each pointed observation by the standard $\chi^2$-maximisation technique. The \texttt{efsearch} task runs recursive fine-searches for periodicities in a time series by folding the data over a range of periods around an estimated period at high resolution. 

The 1s binned light curve was folded with 65536 pulsations around the approximate value of 4.796s obtained from the power density spectrum with a 0.00001s time resolution and 64 phase bins/period. The resultant output is a distribution between the $\chi^2$ value and the spin period of the pulsar that contains a Gaussian peak located at the best fit period which can be measured by fitting a Gaussian model to this pulsation peak. For example, for the $4^{\text{th}}$ time segment, the value of spin period was found to be $4.79534$s. The width of the fitted Gaussian model to the peak is a measure of the error associated with the value of $P_s$.

\begin{figure}
    \centering
    \includegraphics[width=\columnwidth, trim={0 10 0 0}]{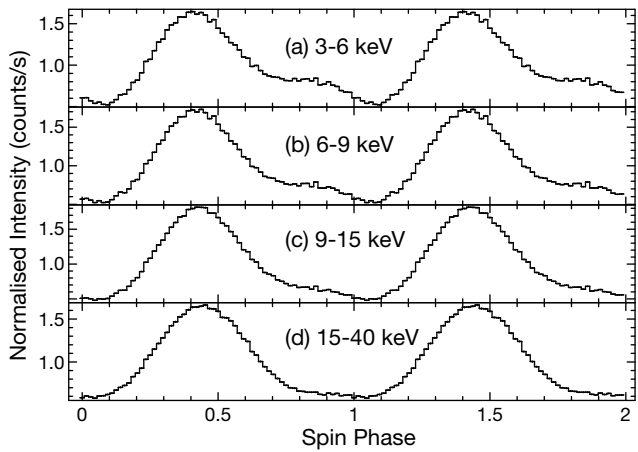}
    \caption{\small Energy resolved pulse profile of Centaurus X-3 in different X-ray energy bands (a) 3.0-6.0 keV, (b) 6.0-9.0 keV, (c) 9.0-15.0 keV and (d) 15.0-40.0 keV obtained by folding and stacking the combined 1s time binned LAXPC10, LAXPC20, LAXPC30 light curve with a pulse period of 4.79534 s. The pulse profile in the figure was created using \texttt{efold} ftool with 64 phase bins per period and a reference time epoch of $T_0=57734.1$ MJD. The task normalises the output profiles for comparison by dividing by the average count rate.}
    \label{fig:energy_resolved}
\end{figure}

Using this exact value of spin period, the folded and stacked pulse profile was generated with the 1s binned light curve using the FTOOL \texttt{efold} by averaging over all consecutive pulses available in each time segment. The profile clearly consists of a prominent, highly-peaked feature followed by a minor shoulder or bump (Fig. \ref{fig:stability}). We refer to these as the primary peak and the secondary inter-pulse, respectively. Further, we resolve this pulse profile into different energy ranges of 3.0-6.0 keV, 6.0-9.0 keV, 9.0-15.0 keV and 15.0-40.0 keV as shown in Fig. \ref{fig:energy_resolved} at a phase resolution of 64 phase bins per period. As the incident flux varies over the observation due to the motion of the pulsar in its binary orbit, the profiles are plotted with normalised intensity in order to discount the effect of such a variation to facilitate comparison. The secondary inter-pulse can be seen clearly in panels (a) and (b) and gradually subsides in panels (c) and (d), leaving just the primary peak to be observed at higher energies. The pulsed fraction of the pulse profiles was computed using the expression $(f_{max}-f_{min})/(f_{max}+f_{min})$ where $f_{max}$ and $f_{min}$ are the values of maximum and minimum intensity of the profile, respectively (Roy {\em et al.} 2020; Yang {\em et al.} 2018).

\subsection{Estimation of Spin and Orbital Parameters}

For an accreting X-ray pulsar, the observed spin period is modulated by the Doppler shift arising out of its orbital motion which suffers smearing when integrated over a long time interval. To obtain the accurate and refined results of the values of systemic spin and orbital parameters, we corrected for this Doppler modulation using phase calibration that assigns a value of spin phase to each detected X-ray photon. The photons were then categorised into 10 spin phase bins for further analysis. 

\begin{figure}[!t]
\includegraphics[width=\columnwidth]{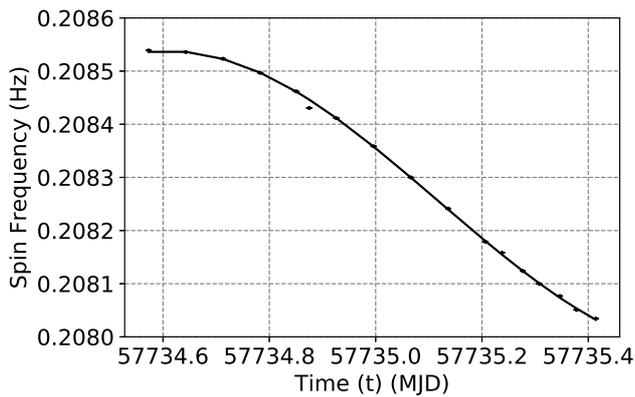}
\caption{The observed spin frequencies have been plotted for all time segments over the full observation time span of MJD 57734.57 - 57735.41. The errors in the measured spin frequencies are smaller than the marker size. The sinusoidal Doppler variation expected to arise from the motion of the pulsar in its binary orbit is over-plotted in a solid black line.}\label{fig:SineFit}
\end{figure}

The Doppler formula for change in frequency $\nu$ is given as,
\begin{equation} \label{eq:Doppler}
    \frac{\Delta \nu}{\nu}=\frac{\text{v}_r}{c}.
\end{equation}
As the orbit of this binary is nearly circular, the radial component of the velocity along the line-of-sight is

\begin{equation} \label{eq:radial_vel}
    \text{v}_r=\text{v}_{b} \sin{\omega_b (t-T_0)},
\end{equation}
where $\text{v}_{b}$ is the orbital velocity, $T_0$ is a reference epoch corresponding to a vanishing $\text{v}_{r}$ and $\omega_b$ is the orbital angular velocity.\\\\
Using Eqs. (\ref{eq:Doppler}) and (\ref{eq:radial_vel}), one can write

\begin{equation} \label{eq:model}
   \nu(t)=\nu_0 \Bigg\{ 1+ \frac{\text{v}_{b}}{c}\sin{\Bigg[ \frac{2\pi (t-T_0)}{P_b} \Bigg]} \Bigg\},
\end{equation}
where $\nu_0$ is the systemic spin frequency of the pulsar observed at the reference epoch $T_0$ and $P_b$ is the orbital period of the binary system. Such a sinusoidal fit to the spin frequencies shown in Fig. \ref{fig:SineFit} was obtained using the Levenberg-Marquardt algorithm, with $\nu_0$, $\text{v}_b$, $T_0$ and $P_b$ as free parameters. 

We integrate Eq. (\ref{eq:model}) to derive the value of phase with time given as,

\begin{equation} \label{eq:phase}
\phi(t)=\nu_0 \Bigg\{ (t-T_0) - \frac{\text{v}_{b} P_b}{2 \pi c} \Bigg[ 1- \cos{\Bigg( \frac{2 \pi (t-T_0)}{P_b} \Bigg) } \Bigg] \Bigg\}.
\end{equation}

\begin{figure}[!t]
\includegraphics[width=\columnwidth]{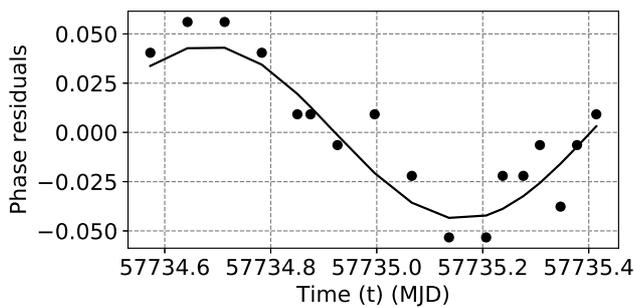}
\caption{The figure shows the residuals in phase for all time segments after preliminary calibration carried out by correcting for the Doppler variation in measured spin frequencies. The residuals show a sinusoidal trend with a periodicity of $\sim P_{b}/2$ typical of a small deviation from the circular orbit approximation. The root-mean-square (rms) variation is 0.033.}
\label{fig:resi_before}
\end{figure}

Applying Eq. (\ref{eq:phase}) to the time stamp of each detection, we assigned a phase value to each event and constructed phase histograms individually for each time segment, thus yielding the corresponding pulse profiles. It can be seen in Fig. \ref{fig:stability} that the profiles do not align exactly. The differences in phases of the pulse profiles were obtained by locating the peak of the pulse profile for each time segment and is shown as a function of time in Fig. \ref{fig:resi_before}. The validity of this method is attributed to the stability of the shape of the pulse as verified in Fig. \ref{fig:stability} which prevents ambiguities in measurements of phase due to pulse shape variations and thus, provides a unique time marker for each of the pulse peaks. A systematic sinusoidal trend is evident, which we fit with a model with a periodicity of $P_b/2$ which is characteristic of a small deviation of the shape of the binary orbit from a circular approximation (Fabbiano \& Schreier 1977). Despite the upper constraint on the orbital eccentricity of Centaurus X-3 being placed at a very small 0.0001 by Raichur \& Paul (2010), we corrected for this effect nevertheless in order to derive precise systemic parameter measurements. 

\begin{figure}[!t]
\includegraphics[width=\columnwidth]{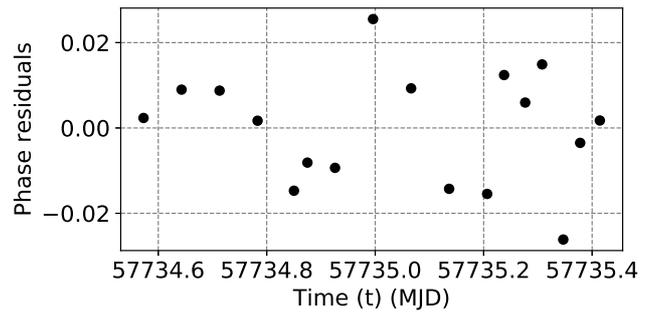}
\caption{The figure shows the residuals in phase for all time segments after fine calibration that corrects for the small deviation from circular orbit approximation to ensure accuracy of results. The residuals are randomly distributed and do not exhibit any systematic trend, which is evident from the slight variation in pulse profile during the observation span. The root-mean-square (rms) variation is 0.013.}
\label{fig:ResidualPhaseCorr}
\end{figure}

\section{Results and Discussions}

\begin{table} 
    \centering 
    \begin{threeparttable}
    \caption{Measured values of the spin and orbital parameters of Centaurus X-3 using AstroSat 2016 observations}
    \begin{tabular}{lc} \topline
      Parameter   &  Value\\ \midline
      \textit{Pulsar Spin} &\\ 
       $P_s$ & $4.80188 \pm 0.000085$  s \\ 
       $T_0$ & $57734.10 \pm 0.01$ MJD\\
       $<\dot{P_s}>|_{1997}^{2016}$  \tnote{a,b} & $-(1.978 \pm 0.015) {\times} 10^{-11} \text{ s/s}$\\ 
        $ | P_s/ \dot{P_s} | $  & $ 7709 \pm 58$ yr \\ \\
       \textit{Binary Orbit} & \\
       $P_b$ &  $2.033 \pm 0.029$ days\\ 
        $<\dot{P_b}>|_{1971}^{2016}$  \tnote{a,c}  & $-(1.19 \pm 0.63) {\times} 10^{-3}$ days/yr  \\
       $\dot{P_b}/P_b $ & $-(5.78 \pm 2.9) {\times} 10^{-4} \text{ yr}^{-1}$ \\
       $\text{v}\sin{i}$ & $410.2 \pm 6.2$  km/s \\
       $a\sin{i}$  & $38.23 \pm 0.58$ lt-s\\\hline
    \end{tabular}
    \label{tab:systemicparam}
    \begin{tablenotes}
    \item[a] These represent average values estimated from the difference of the measured periods at the indicated epochs and dividing them by the time interval
    \item[b] From Burderi {\em et al.} (2000) to this work
    \item[c] From Schreier {\em et al.} (1972b) to this work
    \end{tablenotes}
    \end{threeparttable}
\end{table}


The timing characteristics of the high mass X-ray binary Centaurus X-3 have been presented from AstroSat/LAXPC observations on MJD 57735. The count rate measured during our observation varied between 1187 counts/s and 1356 counts/s. The minimum value is still much larger than the expected drop of an order of magnitude during an eclipse (Giacconi {\em et al.} 1971; Schreier {\em et al.} 1972b). Thus, it was confirmed that an eclipsing event did not occur during the time span of our observation.

A prominent pulse peak with a secondary inter-pulse was seen in the pulse profile in Fig. \ref{fig:stability} which is in agreement with that reported in previous literature (Suchy {\em et al.} 2008; Burderi {\em et al.} 2000). Suchy {\em et al.} (2008) interpret this to be due to a larger fan out of higher energy emission in comparison to the more sharply beamed lower energy X-rays. The pulse profile was observed to have a significant dependence on photon energy as seen in Fig. \ref{fig:energy_resolved}. The pulsed fractions (\%) are $51.8 \pm 0.34, 53.6 \pm 0.37, 58.2 \pm 0.35, 48.4 \pm 0.43$ in the four energy bands in Fig. \ref{fig:energy_resolved}, respectively. A double-peaked pulse profile is seen in the 1-10 keV band which matches the overall pulse profile observed over the full energy band of 3.0-80.0 keV shown in Fig. \ref{fig:stability}. This double-peaked nature gradually evolves into a single-peaked behaviour above 10 keV, consistent with previous reports (Nagase {\em et al.} 1992). Kraus {\em et al.} (1996) have interpreted the shape and the energy-dependence of the pulse profile to point to a distorted magnetic dipole, as also confirmed by Suchy {\em et al.} (2008). 

The measured values of spin and orbital parameters are summarised in Table \ref{tab:systemicparam}. A sinusoidal variation in the measured values of local spin period was observed which is consistent with our expectation from the orbital Doppler effect. The second sinusoidal nature of the phase residuals seen in Fig. \ref{fig:resi_before} was found to have a periodicity of $0.964 \pm 0.011$ days which agrees with $\sim P_b/2$ expected due to a small deviation from the circular orbit approximation (Fig. \ref{fig:ResidualPhaseCorr}). Highly accurate phase calibration was achieved by correcting for both of the above effects as verified by a decrease in the rms of the phase residuals from 0.033 to 0.013. The small randomness remaining in the residuals arises from the slight variation in pulse profile during the observation span (Raichur \& Paul 2010). 

The systemic spin period of the pulsar in Centaurus X-3 was found to be $4.80188 \pm 0.000085$ s at the epoch of our observation. It has decreased from the last reported value of 4.81423 $\pm$ 0.00001 s (Burderi {\em et al.} 2000) indicating that the neutron star is continuing to spin up, which is consistent with its known behaviour. Assuming a constant spin up rate, an estimate of the average spin up timescale over a $\sim$ 20 year stretch from MJD 50507.14 (Burderi {\em et al.} 2000) is found to be $| P_s/\dot{P_s}|=7709 \pm 58 \text{ yr}$, more than twice of $3571.428$ yr reported earlier by Schreier {\em et al.} (1972b). This retardation in the spin up rate may indicate variation in the transfer of angular momentum to the neutron star induced by a corresponding variation in the mass accretion rate (Tsunemi {\em et al.} 1996). A constant spin up rate is indeed an approximation -- variations including glitches have been observed to be present (Bildsten {\em et al.} 1997).

The orbital period was found to be $2.033 \pm 0.029$ days, giving an average value of the orbital decay rate $\dot{P_b}/P_b$ as $-(5.78 \pm 2.9) {\times} 10^{-4} \text{ yr}^{-1}$, with the largest possible time baseline of 45 years available in the literature from MJD 41131.58 (Schreier {\em et al.} 1972b) till this work. Although the significance of our measurement is somewhat limited, this value appears to be substantially higher than that of $-(1.799 \pm 0.002) {\times} 10^{-6} \text{ yr}^{-1}$ reported by Raichur \& Paul (2010). This could indicate short time scale variations in the orbital period as noticed by Kelley {\em et al.} (1983).

The orbital velocity projected along our line-of-sight was found to be $410.2 \pm 6.2$ km/s which is similar to $\text{v}\sin{i}=415.1 \pm 0.4$ km/s reported by Schreier {\em et al.} (1972b). The projected orbital radius $a\sin{i}$ earlier measured to be $ 39.6612 \pm 0.0009$ lt-s by Raichur \& Paul (2010) was found to have decreased to $38.23 \pm 0.58$ lt-s in accordance with the orbital decay of the Centaurus X-3 binary system.

\section{Conclusions}
We have carried out timing analysis of an out-of-eclipse X-ray observation of Centaurus X-3 using AstroSat/LAXPC. We find that the broad-band 3-80 keV pulse profile of Centaurus X-3 has a prominent pulse peak with a secondary inter-pulse which is consistent with previous observations (Suchy {\em et al.} 2008). The systemic spin period, corrected for the orbital motion of the X-ray pulsar around its optical counterpart and the small value of orbital eccentricity, was found to have decreased to $4.80188 \pm 0.000085$ s from $4.81423 \pm 0.00001$ s (Burderi {\em et al.} 2000) which is in agreement with the spin up trend observed in Centaurus X-3. 
The projected semi-major axis and orbital velocity were found to be $38.23 \pm 0.58$ lt-s and $410.2 \pm 6.2$ km/s, respectively. We observe an increase in the spin up timescale by over a factor of 2 to $7709 \pm 58$ yr. This deceleration in the spin up rate of the pulsar could result from a re-proportioning of positive and negative torque action in the innermost regions of the accretion disk.

\section*{Acknowledgements}

The research is based to a significant extent on the results obtained from the AstroSat mission of the Indian Space Research Organisation (ISRO), archived at the Indian Space Science Data Centre (ISSDC). This work has used data from the Large Area X-ray Proportional Counter (LAXPC) detectors developed at TIFR, Mumbai and we thank the LAXPC Payload Operation Centre for verification and release of the data on the ISSDC data archive and providing the requisite software for analysis. We thank the AstroSat Science Support Cell (ASSC) hosted at IUCAA for technical assistance. This research has made use of the High Energy Astrophysics Software obtained through the High Energy Astrophysics Science Archive Research Center (HEASARC) Online Service, provided by the NASA/Goddard Space Flight Center (GSFC), in support of NASA's High Energy Astrophysics Programs. This research has also made use of the NumPy and SciPy packages in Python. Valuable suggestions from an anonymous referee have improved the presentation of this paper.

\def\apj{ApJ}
\def\mnras{MNRAS}
\def\aap{A\&A}
\def\apjl{ApJL}
\def\baas{Bulletin of the AAS}
\def\physrep{PhR}
\def\aapr{A\&AR}
\def\prl{Ph.~Rev.~Lett.}
\def\apjs{ApJS}
\def\pasa{PASA}
\def\pasj{PASJ}
\def\japa{JApA}
\def\nat{Nature}
\def\memsai{MmSAI}
\def\aj{AJ}
\def\aaps{A\&AS}
\def\iaucirc{IAU~Circ.}
\def\sovast{Soviet~Ast.}
\def\apss{Ap\&SS}
\def\procspie{Proc.~SPIE}





\begin{theunbibliography}{}
\vspace{-1.5em}

\bibitem{agrawal2006}
Agrawal, P. C. 2006, Advances in Space Research, 38, 2989

\bibitem{agrawal2017}
Agrawal, P. C., Yadav, J. S., Antia, H. M. {\em et al}. 2017, {\em J. Astrophys. Astr.}, 38, 30

\bibitem{antia}
Antia, H. M., Yadav, J. S., Agrawal, P. C. {\em et al}. 2017, ApJS, 231, 10

\bibitem{ash1}
Ash, T. D. C., Reynolds, A. P., Roche, P. {\em et al}. 1999, MNRAS, 307, 357

\bibitem{bhattacharya}
Bhattacharya, D. 2017, {\em J. Astrophys. Astr.}, 38, 51

\bibitem{bildsten}
Bildsten, L., Chakrabarty, D., Chiu, J. {\em et al}. 1997, ApJS, 113, 367

\bibitem{burderi}
Burderi, L., Di Salvo, T., Robba, N. R. {\em et al}. 2000, ApJ, 530, 429

\bibitem{chodil}
Chodil, G., Mark, H., Rodrigues, R. {\em et al}. 1967, Ph. Rev. Lett., 19, 681

\bibitem{day}
Day, C. S. R., Stevens, I. R. 1993, ApJ, 403, 322

\bibitem{fabbiano}
Fabbiano, G., Schreier, E. J. 1977, ApJ, 214, 235

\bibitem{falanga}
Falanga, M., Bozzo, E., Lutovinov, A. {\em et al}. 2015, A\&A, 577, A130

\bibitem{giacconi}
Giacconi, R., Gursky, H., Kellogg, E. {\em et al}. 1971, ApJL, 167, L67

\bibitem{hutchings}
Hutchings, J. B., Cowley, A. P., Crampton, D. {\em et al}. 1979, ApJ, 229, 1079

\bibitem{kelley}
Kelley, R. L., Rappaport, S., Clark, G. W., Petro, L. D. 1983, ApJ, 268, 790

\bibitem{kraus}
Kraus, U., Blum, S., Schulte, J. {\em et al}. 1996, ApJ, 467, 794

\bibitem{krzeminski}
Krzeminski, W. 1974, ApJL, 192, L135

\bibitem{nagase1989}
Nagase, F. 1989, PASJ, 41, 1

\bibitem{nagase1992}
Nagase, F., Corbet, R. H. D., Day, C. S. R. {\em et al}. 1992, ApJ,396, 147

\bibitem{naik}
Naik, S., Paul, B., Ali, Z. 2011, ApJ, 737, 79

\bibitem{paul}
Paul, B. 2017, {\em J. Astrophys. Astr.}, 38, 39

\bibitem{petterson}
Petterson, J. A. 1978, ApJ, 224, 625

\bibitem{raichur}
Raichur, H., Paul, B. 2010, MNRAS, 401, 1532

\bibitem{rawls}
Rawls, M. L., Orosz, J. A., McClintock, J. E. {\em et al}. 2011, ApJ, 730, 25

\bibitem{roy}
Roy, J., Agrawal, P. C., Singari, B. Misra, R. 2020, RAA, 20, 155

\bibitem{santangelo}
Santangelo, M., Benedetti, W., Donati, S. {\em et al}. 1988, Astronomia UAI, 6, 36

\bibitem{schreier1972a}
Schreier, E., Levinson, R., Gursky, H. {\em et al}. 1972a, ApJL, 172, L79

\bibitem{schreier1972b}
Schreier, E., Tananbaum, H., Kellogg, E. {\em et al}. 1972b, in Bulletin of the AAS, Vol. 4, 261

\bibitem{singh}
Singh, K. P., Tandon, S. N., Agrawal, P. C., {\em et al}. 2014, ASTROSAT mission, Society of Photo-Optical Instrumentation Engineers (SPIE) Conference Series, Vol. 9144, 91441S

\bibitem{stella}
Stella, L., Angelini, L. 1992, in Data Analysis in Astronomy, 59

\bibitem{suchy}
Suchy, S., Pottschmidt, K., Wilms, J., {\em et al}. 2008, ApJ, 675, 1487

\bibitem{thompson}
Thompson, T. W. J., Rothschild, R. E. 2009, ApJ, 691, 1744

\bibitem{tjemkes}
Tjemkes, S. A., Zuiderwijk, E. J., van Paradijs, J., 1986, A\&A, 154, 77

\bibitem{tsunemi}
Tsunemi, H., Kitamoto, S., Tamura, K. 1996, ApJ, 456, 316

\bibitem{verbunt}
Verbunt, F., van den Heuvel, E. P. J. 1995, in X-ray Binaries, 457

\bibitem{white}
White, N. E., Swank, J. H. 1982, ApJL, 253, L61

\bibitem{yadav}
Yadav, J. S., Agrawal, P. C., Antia, H. M. {\em et al}. 2016, in SPIE Conf. Ser., Vol. 9905, Space Telescopes and Instrumentation 2016: Ultraviolet to Gamma Ray, ed. J. W. A. den Herder, T. Takahashi, \& M. Bautz, 99051D

\bibitem{yang}
Yang, J., Zezas, A., Coe, M. J. {\em et al}. 2018, MNRAS, 479, L1

\end{theunbibliography}
\end{document}